%% file: 2017IEEEITS.tex
\documentclass[journal]{IEEEtran}

\usepackage{cite}
\usepackage{color}
\usepackage{url}
\usepackage{hyperref}
\usepackage{amsmath}
\usepackage{amssymb}
\usepackage{graphicx}
\usepackage{cleveref}
\usepackage{algorithm}
\usepackage{algpseudocode}
\usepackage{setspace}
\usepackage[caption=false,font=footnotesize]{subfig}

\title{Data-Driven Spatio-Temporal Analysis of Curbside Parking Demand: A Case-Study in Seattle}
\author{Tanner Fiez and Lillian J.~Ratliff
\thanks{T.~Fiez and L.~J.~Ratliff are with the Department
of Electrical Engineering, University of Washington, Seattle,
WA, 98195 USA e-mail:~{\tt\footnotesize $\{$fiezt, ratliffl$\}$@uw.edu}} %
\thanks{The authors have been supported in part by NSF grants CSN-1646912 and
CNS-1634136. T.~Fiez was also supported in part by an NDSEG Fellowship.}%
\thanks{Code to reproduce work available at
    \href{https://github.com/fiezt/spatial-data-analysis}{github.com/fiezt/spatial-data-analysis}}}

\begin{document}

\pagestyle{plain}
\maketitle

\begin{abstract}
\input{sections/abstract}
\end{abstract}

\section{Introduction}
\label{introduction}
\input{sections/introduction}

\section{Spatio-Temporal Characteristics of Demand}
\label{spatial_temporal}
\input{sections/spatio_temporal}

\section{Gaussian Mixture Model}
\label{gmm}
\input{sections/gmm}

\section{Spatial Autocorrelation}    
\label{spatial_autocorrelation}
\input{sections/spatial_autocorrelation}

\section{Experiments \& Results}
\label{results}
\input{sections/results}

\section{Discussion \& Future Work}
\label{discussion}
\input{sections/discussion}

\bibliographystyle{IEEEtran}
\bibliography{bibtex}

\end{document}

%% file: sections/abstract.tex
Due to rapid expansion of urban areas in recent years, management of curbside parking has become increasingly important. To mitigate congestion, while meeting a city's diverse needs, performance-based pricing schemes have received a significant amount of attention. However, several recent studies suggest location, time-of-day, and awareness of policies are the primary factors that drive parking decisions. In light of this, we provide an extensive data-driven study of the spatio-temporal characteristics of curbside parking. This work advances the understanding of where and when to set pricing policies, as well as where to target information and incentives to drivers looking to park. Harnessing data provided by the Seattle Department of Transportation, we develop a Gaussian mixture model based technique to identify zones with similar spatial parking demand as quantified by spatial autocorrelation. In support of this technique, we introduce a metric based on the repeatability of our Gaussian mixture model to investigate temporal consistency.

%% file: sections/introduction.tex
\IEEEPARstart{D}{eveloping} effective parking policy is challenging owing to a city's desire to balance the competing needs of resource consumers (e.g., transit, local business customers, and shared vehicles) with efficient movement of goods and people as well as support of business district vitality. Failing to do so adequately can have unintended consequences, such as added congestion. 

The impact of drivers cruising in search of parking has been well documented in both the research community, and in the media. Cruising for parking is often touted to be a major contributor to congested traffic.  As reported in~\cite[Chapter 1]{shoup:2005aa}, by pricing curbside parking significantly lower than off-street options, cities are creating perverse incentives that result in unintended consequences (i.e.~added congestion simply due to vehicles cruising). Limited supply in high-demand areas exacerbates the issue. Studies have shown that costs associated with parking-related congestion in terms of lost time, excess use of fuel, and increased pollution can be significant (see, e.g.,~\cite{shoup2007gone,levy2010evaluation,zhang2013air, dowling2017much}).

To combat such negative impacts, cities and researchers are examining a number of strategies for better parking resource management. These efforts can predominantly be divided into work focused on off-street parking and sensor management~\cite{geng2013new, griggs2016design, kotb2016iparker, sun2016discriminated, bagula2015design}, performance based pricing strategies~\cite{qian2012optimal, qian:2012aa, zoeter2014new, dowling2017optimizing}, and behavioral models leveraging data~\cite{pierce2013getting, ratliff:2016aa, yang2017turning}. For a comprehensive overview of smart parking solutions see~\cite{lin2017survey}. The work in this paper most closely resembles the latter research thrust. However, contrary to prior work on data-driven behavioral models for parking resource management, we study demand for parking as a function of location instead of price.

To motivate this line of research, we remark that there is mounting evidence supporting the basis that many factors beyond price drive parking decisions. Survey results from Los Angeles (LA) indicated that for those looking to park, proximity to the intended destination is a more important consideration than both the cost of parking and the time spent searching for parking~\cite{glasnapp2014understanding}. On average, the respondents said the maximum distance they would be willing to park from their intended destination was just $3.07$ blocks. Additionally, over half of the respondents parked within one block of their intended destination. In a study conducted in Beijing, nearly $90\%$ of those surveyed said they made their choice of where to park based on proximity to their intended destination, with less than $1\%$ saying low price was the reason \cite{ma2013parking}. In a similar vein as the LA study, nearly $70\%$ of respondents parked within less than a five minute walk to their final destination.

Studies on the price elasticity of parking demand reinforce that price should not be the only control method explored by researchers and municipalities alike. Indeed, an evaluation of the SFpark study\footnote{{\tt sfpark.org}} revealed that while on average price elasticities were negative during the period of study---indicating a decrease in demand following an increase in price or an increase in demand following a decrease in price---elasticities varied immensely with the location, time of day, day of week, and date of price change. Strikingly, the elasticities were often positive, which combined with the previous observations illustrate price is not the the most important factor in decisions~\cite{pierce2013getting}. 

The inadequate awareness of changes in price, and parking policy broadly, is also problematic. In the SFpark study, e.g., driver behavior did not change until there was an increase in marketing and advertisement during the second price adjustment. Moreover, price elasticities stabilized at relatively meager levels following initial volatility, implying that once awareness had become sufficiently high, drivers who were willing to change their behavior had done so \cite{pierce2013getting}. Similar outcomes are also observed in the aforementioned survey in LA which confirmed the awareness of drivers to parking policy and mobile applications for parking is unquestionably low. Of those surveyed only $31$\%, $24$\%, $25$\% were aware of price changes, time of day pricing, and mobile parking applications, respectively \cite{glasnapp2014understanding}. Interestingly, replacing traditional meters with smart meters and smartphone applications only reduces awareness~\cite{carney2013bringing}.

Although the widespread adoption of digital technologies has led to a wealth of new data that could potentially support sophisticated management strategies, cities generally employ simple policies. This is likely due to the fact that policy makers are largely risk-averse and there are high costs associated with making sweeping changes. 

The policies currently in use are not inherently bad, however, as they do benefit from being easy to track and understand. For instance, existing policies customarily utilize static pricing schemes, often with morning and evening rate periods, set uniformly over an extensive number of block-faces which are grouped over arbitrary regions or within existing neighborhood boundaries. While maintaining salient features of the existing policies which make them viable, the approach to selecting where and when to set them can be greatly improved by exploiting newly available data streams.

The obstacles cities face in implementing significant policy changes serve as one of the main motivations of our work. In contrast to prior research, we analyze frequently overlooked factors in parking decisions, such as location and time of day, to propose methods that can improve traditional policies with straightforward modifications. Leveraging available data sources to gain insights into the spatio-temporal characteristics of parking demand, we develop approaches to identify zones and time periods with similar spatial and temporal demand. Such analysis allows for simple, static pricing schemes to be more effective by providing policy makers data-informed suggestions of how groups of block-faces should be partitioned. It also enables novel methods for spatially and temporally redistributing parking demand via, e.g., information dissemination coupled with verifiable incentives. 

Specifically, we show that a Gaussian mixture model (GMM) can be used to identify groups of spatially close block-faces which have a high degree of spatial autocorrelation in observed occupancy patterns. The Bayesian information criterion is employed to select the number of partitions, improving upon the status quo of making this decision either arbitrarily or through heuristics. We supplement the model by providing a method based on the repeatability of the GMM to metricize the consistency of parking demand, and demonstrate through experiments that spatial demand is indeed consistent over time. 

Furthermore, the GMM partitioning is consistent even with price changes. Likewise, while occupancy fluctuations can be significant from season to season, the GMM partition remains consistent. Both of these results reaffirm that location is the primary driver of parking choice. Finally, we remark that the GMM partition is consistent with the location features. For example, block-faces abutting areas with significant tourist attractions are clustered as are those in primarily residential and commercial zones. This is true even in highly mixed-use neighborhoods. In on-going research, we are exploiting identified spatio-temporal features to determine where and when to target information coupled with incentives for affecting parking behavior.

This paper builds on our prior work in~\cite{fiez:2017aa} by significantly increasing the depth of analysis and experimental results. The extensions include, but are not limited to, exploring the effects of price changes and seasonality on both neighborhood-wide occupancy and spatial demand, considering additional methods for designing the spatial weight matrix when evaluating spatial autocorrelation, and examining a broader spectrum of parking areas to test our techniques.

The rest of this paper is organized as follows. In Section \ref{spatial_temporal}, we describe our data sources and the method used to estimate demand and explore the spatio-temporal characteristics of parking demand. We describe our approach using a GMM to identify zones with similar spatial demand and our method to quantify this using spatial autocorrelation in Sections \ref{gmm} and \ref{spatial_autocorrelation}, respectively. In Section \ref{results}, we present the results of our analysis using parking data from the city of Seattle and conclude with a discussion and remarks on future work in Section \ref{discussion}.

%% file: sections/spatio_temporal.tex
We focus our analysis on Seattle, WA and primarily investigate the Belltown neighborhood. Belltown is a rapidly growing mixed use development which blends residential, commercial, and industrial areas. In addition, Belltown has both the highest population density \cite{statatlas} and the most complete coverage of curbside parking of any neighborhood in Seattle.

\begin{figure}[h]
    \centering
    \includegraphics[width=.7\linewidth]{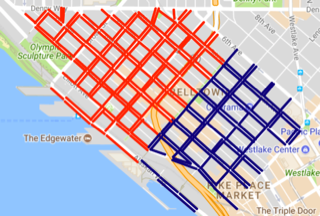}
    \caption{Paid parking in the Belltown neighborhood is divided into the North Zone (red) and the South Zone (blue).}
    \label{fig:belltown_division}
\end{figure}

We provide a map of the paid curbside parking in Belltown in Fig.~\ref{fig:belltown_division}. Parking in the North Zone (red) during $2017$ is \$$1.00$/hr between $8$AM--$11$AM and \$$1.50$/hr between $11$AM--$8$PM with four hour time limits. In the South Zone (blue) parking during $2017$ is \$$2.50$/hr between $8$AM--$5$PM and $5$PM--$8$PM with two and three hour time limits respectively.

\begin{figure*}[htbp]
    \centering
    \includegraphics[width=\linewidth]{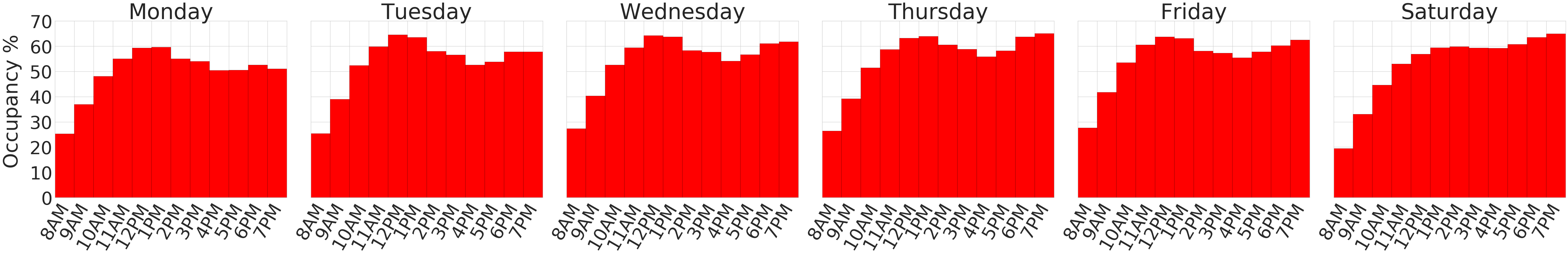}
    \caption{Mean occupancies at active paid parking times in Belltown within Summer 2017.}
    \label{fig:temporal_hetero}
\end{figure*}

\begin{figure*}[htbp]
    \centering
    \hfill\subfloat[][Friday 7PM]{\includegraphics[width=0.4\textwidth]{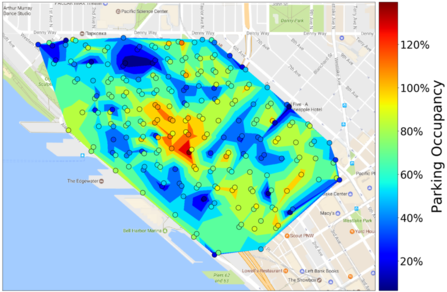}\label{fig:contour1}}\hfill
    \subfloat[][Saturday 11AM]{\includegraphics[width=0.4\textwidth]{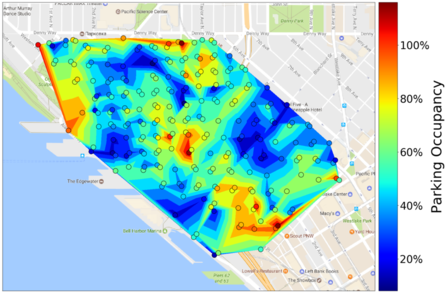}\label{fig:contour2}}\hfill
    \caption{Contours of the mean occupancies in Belltown within Summer 2017 at Friday $7$PM in Fig.~\ref{fig:contour1} and Saturday $11$AM in Fig.~\ref{fig:contour2}. Each scatter point is the midpoint of a block-face.}
    \label{fig:contour}
\end{figure*}

\subsection{Data Sources}
We use paid parking transaction data, block-face supply data, and GPS location data of the block-faces from {June, $2016$--August, $2017$} made available to us via the Seattle Department of Transportation (SDOT)\footnote{These data sources are all publicly available via the open data portal at \href{https://data.seattle.gov}{data.seattle.gov}.}. During this time period there is nearly $14$ million paid parking transactions in Seattle. The paid parking transaction data includes both pay-station and Pay-by-Phone (a mobile app-based payment method) records for each block-face. Paid parking is available Monday--Saturday and typically between $8$AM--$8$PM. The block-face supply data consists of the estimated number of parking spaces for each block-face\footnote{In Seattle, parking spaces are not marked and thus the number of spaces for each block-face is estimated by dividing the length of the legal parking zone into $25$ foot increments.}. The GPS location data of the block-faces includes the latitude and longitude of both ends of a block-face.

\subsection{Demand via Estimated Occupancy}
As a proxy for demand, we use estimated parking occupancy. Due to the prohibitive cost of sensors that would allow for ground truth occupancy to be recorded, this type of data is generally not available. An approach that is more widely applicable is to estimate the parking occupancy using transaction data recorded by smart parking meters. 

We estimate the parking occupancy at a block-face by counting the number of active transactions at the block-face in each minute, and then convert to an occupancy for each minute by dividing by the supply of the block-face. The estimated occupancy at block-face $i$ at time $k$ is given by
\begin{equation}
\text{Occupancy}_i[k] = \frac{\text{Active Transactions}_i[k]}{\text{Supply}_i[k]}.
\end{equation}

Since parking prices generally do not change at any higher frequency than one hour, we aggregate the occupancies up to an hour granularity. The estimated occupancy deviates from the true occupancy for several reasons including: select vehicles are permitted to park for free (disabled permits, government-vehicles, and car-sharing services), vehicles leave before the end of their paid time window, and the estimated supply of a block-face may be inaccurate due to spaces not being marked. These factors can cause the estimated occupancy to be greater than $100\%$, and we clip the maximum estimated occupancy at $150\%$. The estimated occupancy eclipses this limit less than $0.45\%$ of the hourly occupancy instances over all block-faces.  Due to the fact that our analysis is based on the relative relationship between occupancies and assuming that the error in estimating the occupancy has nearly the same effect on each block-face, using the estimated occupancy has a negligible affect on our analysis.

\begin{figure*}[t!]
    \centering
    \subfloat[][Summer 2016 to Fall 2016.]{\includegraphics[width=0.47\linewidth]{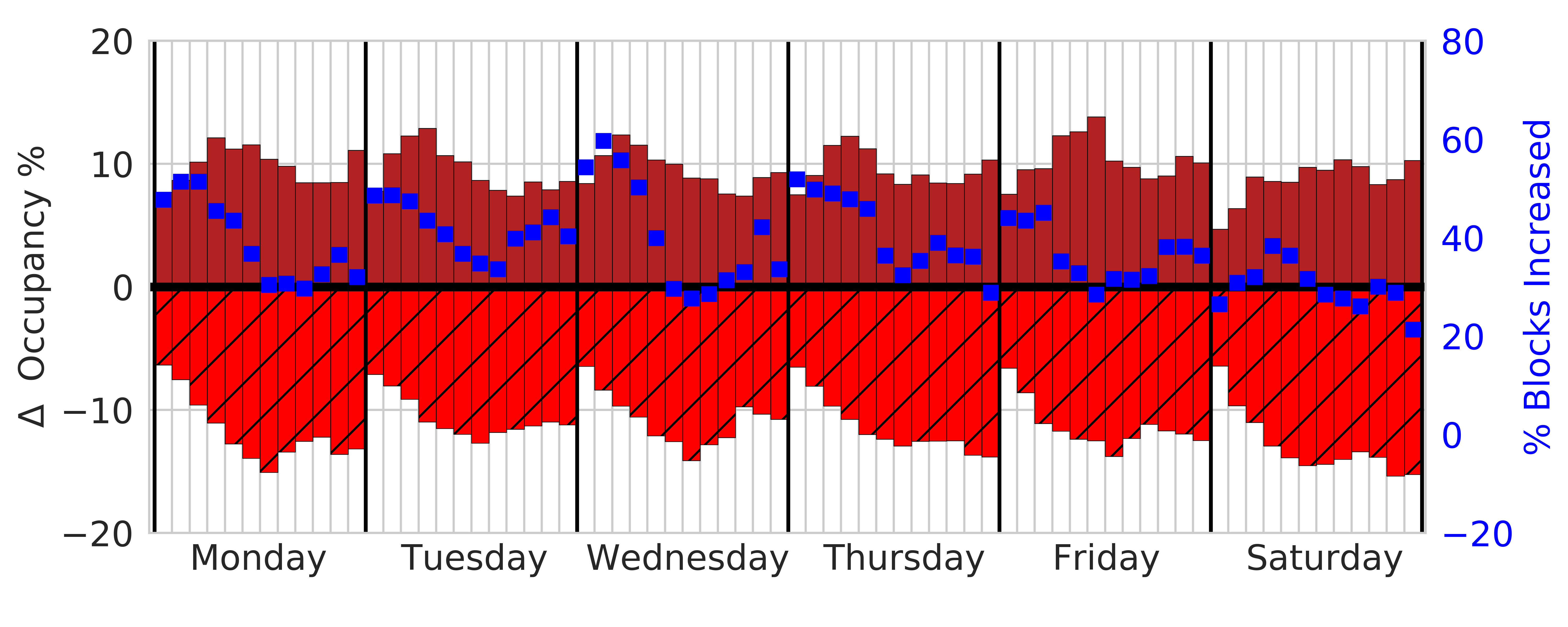}\label{fig:fall_winter}}\hfill
    \subfloat[][Fall 2016 to Winter 2017.]{\includegraphics[width=0.47\linewidth]{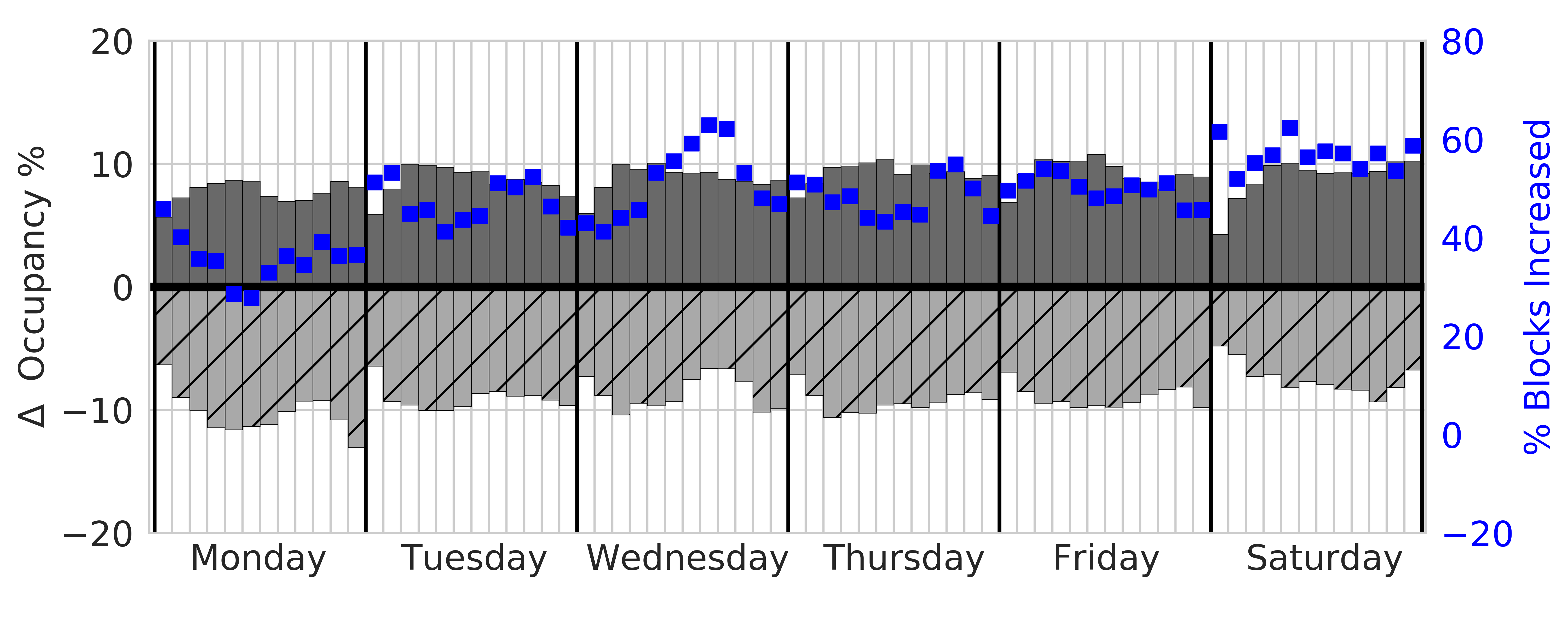}\label{fig:winter_fall}}\newline
    \subfloat[][Winter 2017 to Spring 2017.]{\includegraphics[width=0.47\linewidth]{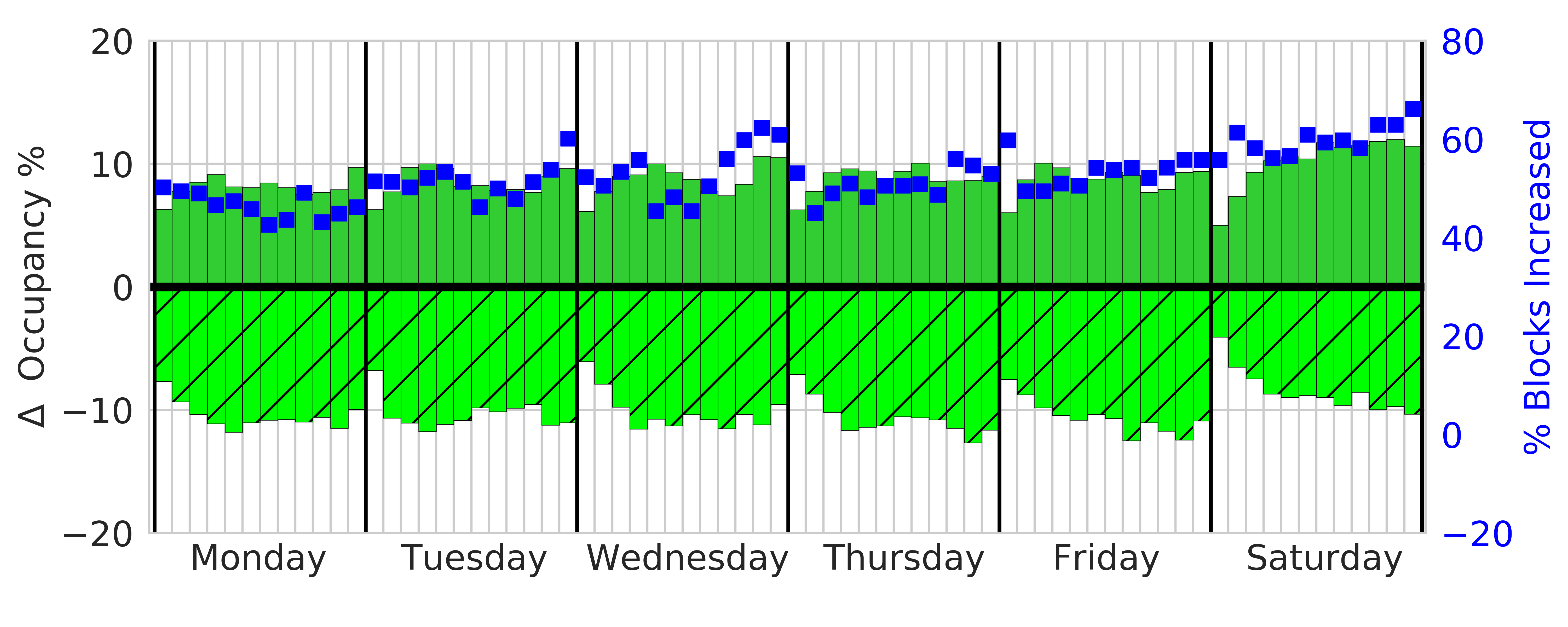}\label{fig:spring_winter}}\hfill
    \subfloat[][Spring 2017 to Summer 2017.]{\includegraphics[width=0.47\linewidth]{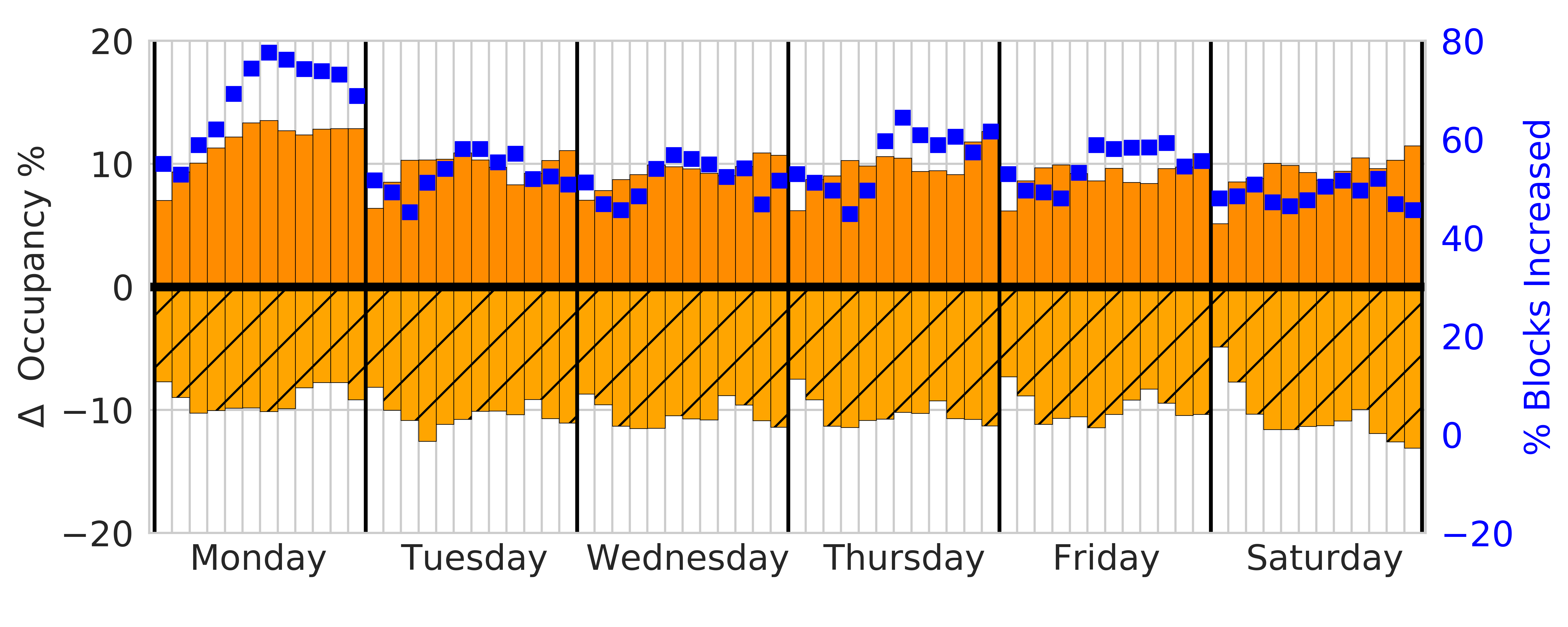}\label{fig:summer_spring}}
    \caption{Seasonality effects on parking demand in Belltown. Each bar indicates an hour of the active paid parking times in a day between 8AM--8PM. The bars above the $x$-axis indicate the mean percentage increase in occupancy at each hour of available paid parking for the block-faces that saw increased occupancy between seasons. The hatched bars below the $x$-axis indicate the mean percentage decrease in occupancy at each hour of available paid parking for the block-faces that saw decreased occupancy between seasons. The blue squares indicate the percentage of block-faces that increased in occupancy between seasons.}
    \label{fig:seasonality}
\end{figure*}

\subsection{Temporal characteristics}
In Fig.~\ref{fig:temporal_hetero} the mean occupancy profiles averaged over the entire Belltown neighborhood are plotted at each hour that paid parking is available. Occupancy profiles for Monday--Friday are very similar, with the exception of Monday having slightly lower occupancy, while the occupancy profile for Saturday follows a different trend. During weekdays, occupancy increases from the start of paid parking until demand peaks near lunch time. It then decreases during the afternoon, until there is another peak during the evening hours when people tend to have dinner. In contrast to weekdays, on Saturday the demand for parking nearly continuously increases throughout the day. These observations highlight that a reasonable parking policy may use unique weekday and weekend pricing schemes, and policies must consider the temporal characteristics of demand that may be driven by neighborhood features such as the presence of certain business types. It is also worth noting that the occupancies are often near the utilization range of $60\%$--$80\%$ that many cities uniformly target independent of local neighborhood characteristics.

\subsection{Spatial characteristics}
We find that spatial demand is heterogeneous and is connected to the temporal characteristics of demand in Belltown. Furthermore, the spatial demand characteristics that we observe are often easily explained. Fig.~\ref{fig:contour} provides a motivating example of these observations. In Fig.~\ref{fig:contour1}, the occupancy at $7$PM on Friday is nearly uniformly distributed throughout Belltown, with the exception that there is an area of much higher occupancy in the center of the neighborhood, which happens to have a high concentration of bars and restaurants that we conjecture drive demand. Interestingly, this area also appears to be up against the divide of the North and South paid parking zones---denoted by red and blue block-faces respectively in Fig.~\ref{fig:belltown_division}---which have a $\$1.00$/hr price difference at this time. One could posit that an improved division of paid parking zones could reduce the congestion in this area by dispersing some of the occupancy into what is now the edge of the South Zone. 

In Fig.~\ref{fig:contour2}, the occupancy at $11$AM on Saturday has a more diverse distribution, but most importantly the areas of high occupancy---with the exception of just a few block-faces---are located in very different locations. The source of the high occupancy areas is immediately clear, as the top and bottom of the neighborhood are the closest parking to some of the most famous weekend tourist attractions in Seattle. Just above the top of the neighborhood is the Space Needle, and just below the bottom of the neighborhood is Pike Place Market. This example highlights one of the key problems we seek to address in this paper: parking policies with uniform pricing schemes in arbitrary zones ignore important properties of spatial demand which reduces their effectiveness.

\subsection{Effects of Seasonality}\label{season}
Parking demand does indeed exhibit fluctuation between seasons. In Fig.~\ref{fig:seasonality} we explore how demand changes between seasons in detail. We find that on average the occupancy of block-faces increases or decreases by approximately $10\%$ between all seasons. Yet, the percentage of block-faces which increase between seasons is heavily dependent on which seasons are being transitioned to and from. From Summer 2016 (Jun.--Aug., 2016) to Fall 2016 (Sept.--Nov., 2016), and similarly between Fall 2016 to Winter 2017 (Dec., 2016--Feb., 2017), at the majority of paid parking times more block-faces decrease in occupancy than do increase. Intuitively, this makes sense. During the day people follow their regular routines---e.g., parking for work---while in the evening, shorter days and worse weather have the effect of causing people to become less likely to go out to businesses and participate in activities. 

There is an analogous trend between Winter to Spring 2017 (Mar.--May, 2017) and Spring to Summer 2017 (Jun.--Aug., 2017). Between these seasons at most of the paid parking times, more block-faces increase in occupancy than decrease. During the morning hours when people follow their normal routines, the percentage of block-faces whose occupancy increases is often at or just above $50\%$. At times in which demand may be driven more by businesses such as the middle of the day near lunch and in the evening, a higher percentage of block-faces increase in occupancy. The intriguing property of these observations is that despite variations between seasons, the way occupancy is distributed, i.e.~the spatial demand, does not vary significantly. This confirms that static policy schemes considering location can be robust to the effects of seasonality. We explore this further in Section~\ref{results}.

\subsection{Effects of Price Changes}\label{price}
In Seattle, parking prices change once per year after an annual parking study is conducted. In July, 2016 the price to park in the North Zone of Belltown (red block-faces in Fig.~\ref{fig:belltown_division}) decreased from \$$1.50$/hr to \$$1.00$/hr in $8$AM--$11$AM. To investigate the impact the price change had on parking behavior we examined the month before the price change, June, $2016$, and the month one year following the price change, June, $2017$.

\begin{figure}[h]
    \centering
    \subfloat[][North Zone in Belltown.]{\includegraphics[width=0.45\textwidth]{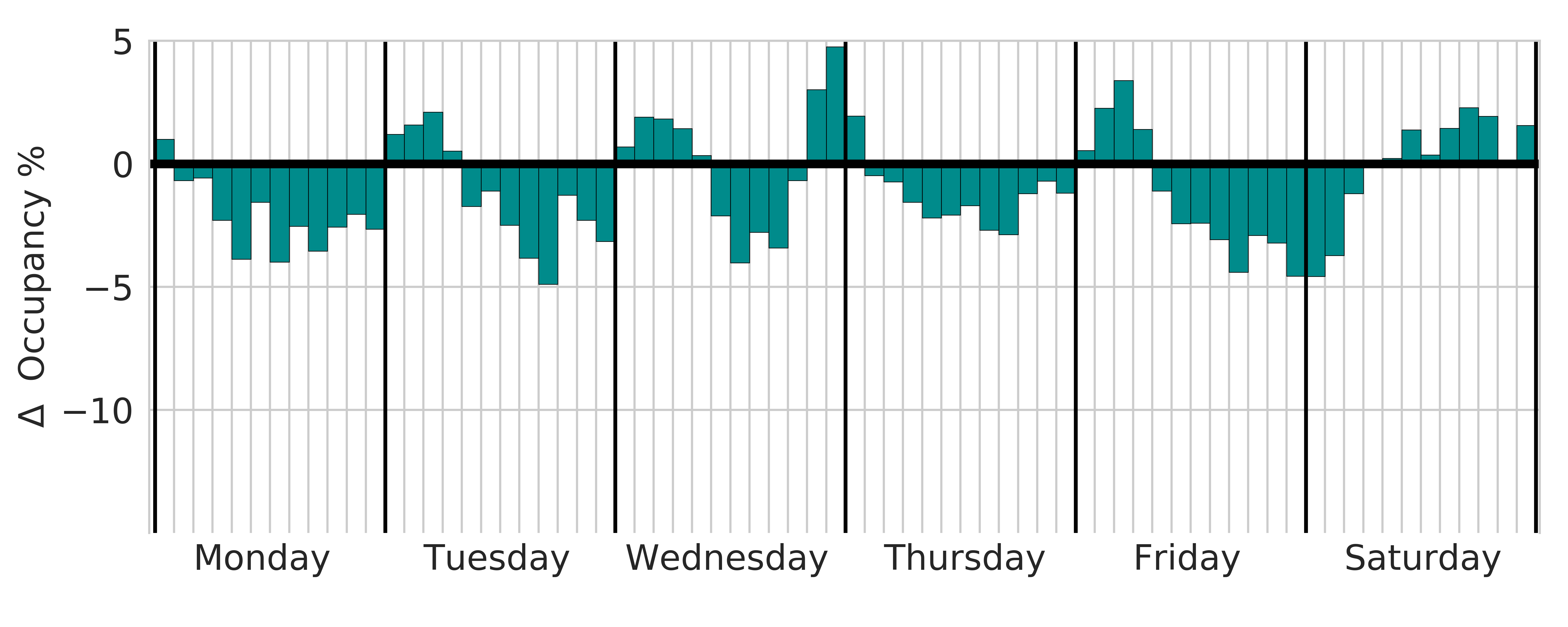}\label{fig:north}}\hfill
    \subfloat[][South Zone in Belltown.]{\includegraphics[width=0.45\textwidth]{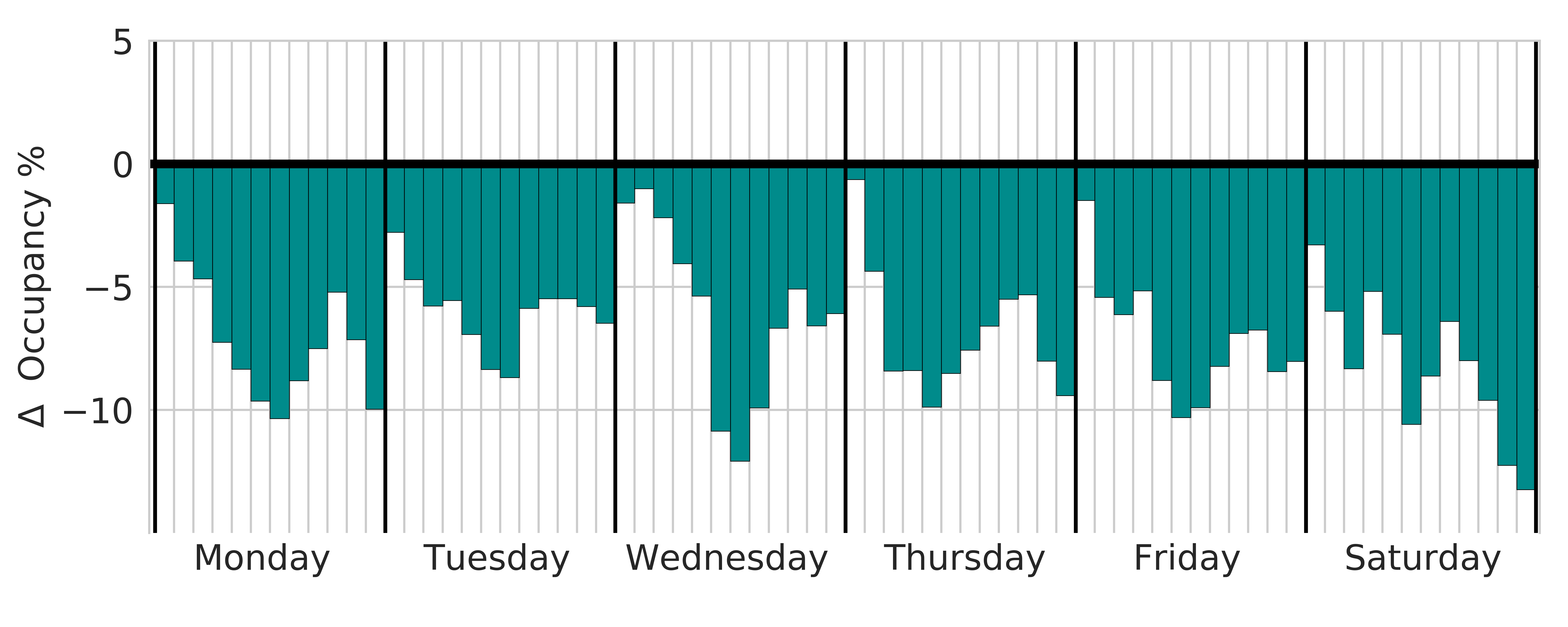}\label{fig:south}}
    \caption{Change in occupancy in the North Zone (Fig.~\ref{fig:north}) and the South Zone (Fig.~\ref{fig:south}) from before (June, 2016) to after (June, 2017) the price change in Belltown.}
    \label{fig:price_diff}
\end{figure}

Fig.~\ref{fig:price_diff} shows the relative change in occupancy during this time period in the North Zone (Fig.~\ref{fig:north}) and the South Zone (Fig.~\ref{fig:south}). It is interesting that in the North zone the times that do see an increase in occupancy primarily occur within the time interval in which the price was decreased. The change is rather insignificant though, and while the South Zone saw decreased occupancy at all times, the smallest decreases in occupancy also occur within the time interval of the price change in the North Zone indicating that price was likely not the factor causing the change in behavior. This confirms much of the prior work showing elasticity to price is mixed and, furthermore, it re-affirms that pricing alone may have limited effect on driver behavior. The variation in occupancy over the entire year on average for block-faces is approximately $20\%$ where between seasons we observed approximately a variation of $10\%$ on average in Fig.~\ref{fig:seasonality}. Commensurate with variation between seasons however, we find the spatial demand does not vary significantly before and after the price change. 

Historically in Seattle utilization of curbside parking has been consistently increasing and the decline in demand is a new trend\footnote{Personal communication with SDOT.}. A possible cause of this may be the recently expanded rail system and increased use of ride-sharing services (Uber, Lyft) as well as bike sharing services (Spin, Lime Bike, ofo).

%% file: sections/gmm.tex
We model parking demand with a GMM by using it as an unsupervised clustering method to find zones and groups of block-faces within them that are spatially close and have similar demand. 
This technique enables us to:
\begin{enumerate}
\item Draw new inferences about parking demand and its spatio-temporal characteristics.
\item Use data to make informed decisions about zones and time periods in
    which static, uniform pricing schemes would be more effective than if chosen arbitrarily.
\item Consider identified zones as groups of users with similar preferences,
    facilitating targeted information and incentives.
\end{enumerate}

\subsection{Model Description}
The GMM is a probabilistic method to model a distribution of data with a mixture of multivariate Gaussian distributions, each with a mean vector $\mu_j$ and covariance matrix $\Sigma_j$. The probability distribution of the GMM with $k$ mixture components is given by 
\begin{equation}\textstyle
p(x_i|\pi, \mu, \Sigma) = \sum_{j=1}^k\pi_j \mathcal{N}(x_i|\mu_j, \Sigma_j).
\end{equation}
\indent
We consider each sample of our dataset to be a vector $x_i \in \mathbb{R}^3$, containing spatial and demand features at a given time for a block-face as 
\begin{equation}
x_i = \begin{bmatrix}x_{i,\text{latitude}} & x_{i,\text{longitude}} & x_{i,\text{occupancy}}\end{bmatrix}.
\end{equation}
Thus the complete dataset is given by the matrix of the $n$ samples stacked as $x = \begin{bmatrix}x_1 & \cdots & x_n\end{bmatrix}^T$. In our implementation we normalize features column-wise to be in $[0, 1]$. A motivation for the features we choose is their simplicity---as they exactly capture the spatial demand aspects of the data we have---and they work to tradeoff grouping block-faces which are close and block-faces which have similar demand. It is also possible that these features capture unobserved information and characteristics which we do not include that may guide decisions such as the price, type of area (residential, commercial, industrial), type of parking (parallel, angeled), etc. 

Let us now introduce a vector of $n$ indicator variables $z = \begin{bmatrix}z_1 & \cdots & z_n\end{bmatrix}$ as the latent component labels for samples. The prior on the probability of a sample belonging to a mixture component can then be expressed as
\begin{equation}
p(z_i = j) = \pi_j.
\end{equation}
The parameter $\pi$ must satisfy the restrictions that $\pi_j \in [0, 1]$ and $\sum_{j=1}^k\pi_j = 1.$ 

The likelihood of a sample belonging to a mixture component is given by 
\begin{equation}
p(x_i|z_i = j) = \mathcal{N}(x_i|\mu_j, \Sigma_j),
\end{equation}
where the multivariate Gaussian distribution is
\begin{equation}
\textstyle\mathcal{N}(x_i|\mu_j, \Sigma_j) = \frac{\text{exp} \ [{-\frac{1}{2}(x_i - \mu_j)^T\Sigma_j^{-1}(x_i - \mu_j)}]}{(2 \pi)^{\frac{3}{2}}|\Sigma_j|^{\frac{1}{2}}}.
\end{equation}
We make the common assumption that the features are conditionally independent given the component, i.e. each covariance matrix $\Sigma_j$ is diagonal. This modeling decision is also motivated by our application. When selecting zones to design policies or to target information, simple boundaries are more feasible.

The objective function of the GMM is the log likelihood of the data given by
\begin{equation}\label{eq:LL}\textstyle
\text{LL} \triangleq \log \ p(x|\pi, \mu, \Sigma) = \sum_{i=1}^n \log \sum_{j=1}^k\pi_j \mathcal{N}(x_i|\mu_j, \Sigma_j).
\end{equation}
We employ the expectation-maximization (EM) algorithm \cite{dempster1977maximum}
to optimize for this objective. The EM algorithm, given in Algorithm 1, consists
of an initialization of the unknown parameters and two steps, the expectation
step (E-step) and the maximization step (M-step), which are repeated until convergence. The convergence criteria we use is to terminate the algorithm when the change in the log likelihood between iterations, which is ensured to be positive 
since the log likelihood is guaranteed to increase at each iteration of the EM algorithm \cite{wu1983convergence, boyles1983convergence}, is smaller than a parameter $\epsilon$.  
\begin{equation}
\Delta \text{LL} \triangleq \text{LL}^i - \text{LL}^{i-1} \leq \epsilon.
\end{equation}
In the E-step, the expected values of the unobserved component labels given the current parameter values are computed. These are the posterior probabilities and are sometimes referred to as the responsibility that component $j$ takes for data point $i$ \cite{Murphy:2012:MLP:2380985}. Formally, we will denote this term as 
\begin{equation}
r_{i,j} \triangleq p(z_i = j|x_i, \pi_j, \mu_j, \Sigma_j).
\end{equation}
In the M-step, the parameter values are updated to maximize the log likelihood. 

Once the convergence criteria is met, we make hard assignments of each sample $x_i$ to the component label $j$ which maximizes the responsibility $r_{i,j}$---that is,
\begin{equation}\label{eq:assign}\textstyle
z^*_i = \arg\max_j \ r_{i,j}.
\end{equation}
\indent
The objective function is non-convex, which only guarantees that we find local minima. Hence, we run the algorithm for several random initializations and retain the model from the iteration that resulted in the highest log likelihood.

\subsection{Model Selection}\label{model_selection}
The model selection problem for a GMM is to determine the number of mixture components to use in the model. We leverage the Bayesian information criterion (BIC) \cite{schwarz1978estimating} for this purpose. The BIC for our problem is given by 
\begin{equation}
\text{BIC} = -2\text{LL} + \log(n)\nu,
\end{equation}
where LL is the maximized objective defined in (\ref{eq:LL}), $n$ is the number of samples in the dataset, and $\nu$ is the number of degrees of freedom in the GMM. 

The number of degrees of freedom for a model containing $k$ components with $d$ dimensional data is given by ${\nu = k \cdot (2d + 1)}$. Each component contributes $d$ degrees of freedom from both the mean vector and the diagonal covariance matrix, as well as a final degree of freedom from the component prior. 

To determine the number of components to use in our experiments, we performed a search over the value of $k$ averaging the BIC of the GMMs learned on each day of the week and hour of the day combination in our dataset using the mean occupancies at these instances. We then selected the value of $k$ which minimized the mean BIC.

\begin{algorithm}[t]
  \caption{EM Algorithm for GMM}\label{alg:emgmm}
  \begin{spacing}{0.8}
  \begin{algorithmic}[1]
    \Procedure{EM}{$x$}\Comment{$x$: normalized feature matrix}
    \For{each initialization}
    \While{$\Delta \text{LL} > \epsilon$}
    \For{each sample $x_i$}
    \For{$j$ in $\{1,\ldots,k\}$} \Comment{E-Step}
        \State $$r_{i,j} = \frac{\pi_j\mathcal{N}(x_i|\mu_j,
        \Sigma_j)}{\sum_{j^{'}}^k\pi_{j^{'}}\mathcal{N}(x_i|\mu_{j'},
        \Sigma_{j'})}$$ 
        \EndFor
    \EndFor
    \For{each component $\pi_j, \mu_j, \Sigma_j$}\Comment{M-Step}
            \State $$\pi_j =\textstyle \frac{1}{n}\sum_{i=1}^nr_{i,j}$$
            \State $$\mu_j = \frac{\sum_{i=1}^nx_ir_{i,j}}{\sum_{i=1}^n r_{i,j}}$$
            \State $$\Sigma_j = \frac{\sum_{i=1}^n(x_i - \mu_j)^T(x_i -
            \mu_j)r_{i,j}}{\sum_{i=1}^n r_{i,j}}$$
    \EndFor
        \EndWhile\label{euclidendwhile}
        \State $\text{Store maximized} \ \text{LL} \ \text{for the initialization}$
        \State \text{Store sample assignments using (\ref{eq:assign})}
        \EndFor
        \EndProcedure
  \end{algorithmic}
  \end{spacing}
\end{algorithm}

\subsection{Consistency Metric}
An important question with respect to the spatio-temporal properties of parking demand is the consistency of the demand. In other words, we want to quantify how similar demand is from week to week for a given day of week and time of day. Pricing schemes, as well as targeted information and incentive campaigns, can be constructed much more effectively if there is an understanding that without changes in policy or to the system demand characteristics will remain the same. In particular, given that in Sections \ref{season} and \ref{price} we observed that there is a non-trivial variation in the demand at block-faces through time we want to find how consistently spatial demand is distributed.

We propose a method to metricize the consistency of demand based on the repeatability of our GMM approach. Using our dataset, the procedure to determine the consistency metric value at a day of the week and hour of the day is as follows: 
\begin{enumerate}
\item For the chosen day of the week and hour of the day, select a specific date and learn a GMM using the occupancy data at this instance.
\item Assign component labels to each block-face for all other instances with the same day of the week and hour of the day in the dataset using the learned model.
\item Determine the percentage of block-faces which were assigned to the same component as they were in the original GMM that was learned. 
\item Repeat (1)--(3) switching the date on which the GMM is learned, and then average over the percentages computed at each iteration.
\end{enumerate}
We explore this method and discuss the results in Section \ref{results}.

%% file: sections/spatial_autocorrelation.tex
We have claimed that policies which use uniform pricing schemes over arbitrary zones and time periods can be ineffective because they ignore important spatio-temporal characteristics. Intuitively, this claim is to say that if demand through space and time is not homogeneous, then it would be erroneous to set prices uniformly through space and time.

To this end, we use a standard measure of spatial autocorrelation---Moran's $I$ \cite{moran1950notes}---to quantify the degree of spatial homogeneity or heterogeneity present in the demand. Moran's $I$ is defined as 
\begin{equation}\textstyle
I = \frac{N}{\sum_{i}\sum_j w_{i,j}}\frac{\sum_i\sum_jw_{ij}(o_i - \bar{o})(o_j - \bar{o})}{\sum_i(o_i - \bar{o})^2},
\end{equation}
where for our problem $N$ denotes the number of block-faces, $o_i$ denotes the occupancy for block-face $i$, $\bar{o}$ denotes the mean occupancy over all block-faces, and $W=(w_{i,j})_{i,j=1}^{N}$ is a matrix of spatial weights with zeros along the diagonal. 

Values of $I$ range from $-1$ (indicating perfect dispersion) to $1$ (indicating perfect clustering of similar values). The Moran's $I$ value can be used to find a $z-$score and then a $p-$value to determine whether the null hypothesis, that the data is randomly disbursed, can be rejected. 

We are interested in assessing the spatial autocorrelation locally and globally in a neighborhood or region, within currently designated paid parking areas, and within the zones we find with the GMM. The spatial weight matrix $W$ can be designed to evaluate each of these objectives. In particular, we evaluate each method by determining whether the $p$-values are significant using a two-sided $p$-value with a significance measure of $.01$. We report the percentage of instances in our data set---each instance given by the occupancy at a date and time---that are significant. In Section \ref{results} we will delve into the spatial autocorrelation results for each method of creating the spatial weight matrix.

\subsection{Assessing Local and Global Spatial Autocorrelation}\label{local}
In Section \ref{spatial_temporal} we discussed and demonstrated that parking demand displays spatial heterogeneity. A logical follow up question to this observation is whether there is at least local spatial homogeneity. If this were the case, it would imply that it could be possible to find groups of block-faces where there is spatial homogeneity. To evaluate this objective we create the weight matrix by setting values of $w_{ij}$ to $1$ if block-face $j$ is one of the $k$ nearest neighbors to block-face $i$ and $0$ if it is not. We experiment using a range of values for $k$. 

In order to take a more global view of the spatial demand we also experiment creating the weight matrix by using a distance based metric. That is, we set values of $w_{ij}$ to be the Euclidean distance between block-face $i$ and block-face $j$ in terms of the GPS coordinates, normalized between $0$ and $1$ with weight $1$ given to the closest block-face $j$ from block-face $i$ and weight $0$ given to the furthest block-face $j$ from block-face $i$.

\subsection{Assessing Spatial Autocorrelation in Current Zones}\label{area}
We are also interested in the spatial autocorrelation within the currently designated paid parking zones by the city of Seattle. This will help us appraise current policies and provide a means to make comparisons with our method of selecting paid parking zones. To measure the spatial autocorrelation within the current zones we create the weight matrix by setting values of $w_{ij}$ to $1$ if block-faces $i$ and $j$ are in the same parking zone and $0$ if they are not. 

We also investigate a distance based metric within the paid parking zones in a similar manner as described in Section~\ref{local}. In this method we set values of $w_{ij}$ to be the Euclidean distance between block-face $i$ and block-face $j$ in terms of the GPS coordinates, normalized between $0$ and $1$ with weight $1$ given to the closest block-face $j$ from block-face $i$ in the same paid parking zone and weight $0$ given to the furthest block-face $j$ from block-face $i$ in the same paid parking zone. Block-faces in different paid parking zones are given a weight of $0$.

\subsection{Assessing Spatial Autocorrelation in GMM Components}\label{mixture}
One of the aims of the GMM approach is to identify groups of block-faces that are spatially close and have similar demand. This can also be interpreted as finding zones where there is spatial homogeneity. To gauge our success in doing so, and to justify considering the zones as groups of users with similar preferences, we can again use spatial autocorrelation. In this setting we create the weight matrix by setting values of $w_{ij}$ to $1$ when block-faces $i$ and $j$ are in the same GMM component and $0$ when they are not. 

We also use a distance based metric for this objective. We set values of $w_{ij}$ to be the Euclidean distance between block-face $i$ and block-face $j$ in terms of the GPS coordinates, normalized between $0$ and $1$ with weight $1$ given to the closest block-face $j$ from block-face $i$ assigned to the same mixture component and weight $0$ given to the furthest block-face $j$ from block-face $i$ assigned to the same mixture component. Block-faces assigned to different mixture components are given a weight of $0$.

%% file: sections/results.tex
\begin{figure*}[t!]
    \centering
    \subfloat[][Friday 7PM]{\includegraphics[width=0.33\textwidth]{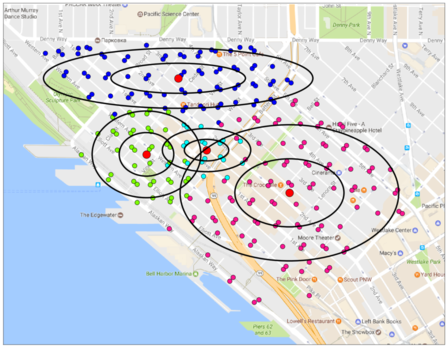}\label{fig:mixture1}}\hfill
    \subfloat[][Saturday 11AM]{\includegraphics[width=0.33\textwidth]{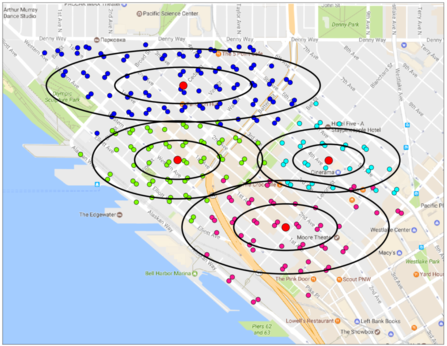}\label{fig:mixture2}}\hfill
    \subfloat[][Wednesday 10AM]{\includegraphics[width=0.33\textwidth]{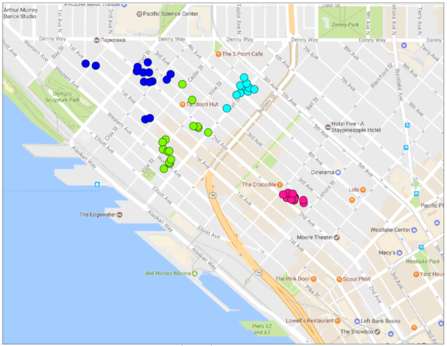}\label{fig:centroids}}
    \caption{GMMs with four components learned using the mean occupancies in
        Belltown within Summer 2017 at Friday $7$PM in
        Fig.~\ref{fig:mixture1} and at Saturday $11$AM in
        Fig.~\ref{fig:mixture2}. Block-faces are colored by the mixture
        component label. The ellipses indicate the first and second standard
        deviations of the components GPS coordinates, with the red scatter
        points indicating the centers of the components GPS coordinates.
        Fig.~\ref{fig:centroids} shows clustering of the spatial centers of the
    mixture components from the GMMs learned in Belltown on all Wednesdays at
10AM in Summer 2017.}
    \label{fig:mixture_centroids}
\end{figure*} 

\begin{table*}[t]
  \centering
      \caption{Consistency metric results for Belltown in active paid parking times for Summer 2017. Units are in percentage.}
  \label{table:results_summer}
  {\small     \begin{tabular}{|c||c|c|c|c|c|c|c|c|c|c|c|c||c|}
    \hline
    Day\textbackslash Time & 8AM & 9AM & 10AM & 11AM & 12PM & 1PM & 2PM & 3PM & 4PM & 5PM & 6PM & 7PM & \textbf{Daily} \\
    \hline\hline 
    M & 70.4 & 82.3 & 88.2 & 91.2 & 90.4 & 88.6 & 87.2 & 87.0 & 87.1 & 86.5 & 85.9 & 85.9 & \textbf{85.9} \\ 
    \hline
    Tu & 70.5 & 81.4 & 89.6 & 91.0 & 89.0 & 90.9 & 89.0 & 86.6 & 84.4 & 85.6 & 88.0 & 85.9 & \textbf{86.0} \\
    \hline
    W & 76.1 & 85.8 & 90.7 & 91.4 & 90.5 & 89.9 & 87.8 & 86.9 & 86.2 & 88.3 & 86.5 & 87.5 & \textbf{87.3} \\
    \hline
    Th & 72.0 & 85.5 & 89.3 & 91.5 & 88.0 & 90.1 & 90.1 & 89.5 & 87.3 & 88.1 & 88.2 & 86.8 & \textbf{87.2} \\
    \hline
    F & 76.8 & 87.0 & 89.0 & 89.7 & 89.0 & 88.4 & 89.2 & 89.1 & 85.8 & 86.5 & 86.5 & 85.2 & \textbf{86.8} \\
    \hline
    S & 66.7 & 75.8 & 78.0 & 86.2 & 85.3 & 83.4 & 86.3 & 86.6 & 86.2 & 85.4 & 85.0 & 84.6 & \textbf{82.5} \\
    \hline\hline
    \textbf{Hourly} &
    \textbf{72.1} & \textbf{83.0} & \textbf{87.5} & \textbf{90.2} & \textbf{88.7} & \textbf{88.5} & \textbf{88.3} & \textbf{87.6} & \textbf{86.2} & \textbf{86.7} & \textbf{86.7} & \textbf{86.0} & \\
    \hline
  \end{tabular}

}
\end{table*}

\begin{table}[t]
  \centering
  \caption{Mean daily consistency metric results for Belltown in each
      season/year analyzed. Units are in percentage.}
  \label{table:results_mean}
  {\small      \begin{tabular}{|c||c|}
    \hline
    \textbf{ Season/Year} & \textbf{Mean Daily Consistency}\\
\hline\hline
      Summer/2016 & 86.6\\\hline
     Fall/2016 & 83.9\\\hline
     Winter/2017 & 84.9\\\hline
     Spring/2017 & 85.8\\\hline
     Summer/2017 & 85.9 \\
    \hline
\end{tabular}}
\end{table}
%
%
We now explore the application of our GMM approach and provide analysis to gain
insights into the spatio-temporal characteristics and consistency of parking
demand.

\subsection{Modeling Belltown with a GMM}
In Fig. \ref{fig:contour}, we illustrate the mean spatial demand in Belltown
within Summer 2017 at Friday $7$PM and Saturday $11$AM. Figs.~\ref{fig:mixture1} and~\ref{fig:mixture2} provide an example use of our GMM approach using the same data. It is clear that we are able to find separable zones in which block-faces spatially close are included in the same mixture components. This is important due to the fact that while there may be spatial heterogeneity in Belltown, we are able to find zones in which block-faces have similar demand thereby validating that zone based pricing is viable. 

The example depicted in Figs.~\ref{fig:mixture1} and~\ref{fig:mixture2} also indicates that the model we learn is related to the day of the week and time of day. The model we learn for Friday night, e.g., is very different from the model we learn for Saturday morning, asserting that the spatial component of demand depends on the temporal component. Consequently, in design of pricing policies and information schemes, the questions of where and when to designate them should be considered together. We discuss further spatio-temporal insights in Section~\ref{insights}.

An interesting result of our analysis in Belltown is the model selection problem. Using the model selection criterion described in Section \ref{model_selection}, we find four mixture components---corresponding to four paid parking zones---to be optimal. At present, Belltown has just two paid parking zones in place. This may play a significant factor in why we can improve on the existing policy design method of using heuristics to set paid zone boundaries. We will explore the results quantifying performance in Section~\ref{auto_results}.

\begin{figure*}[t!]
    \centering
    \subfloat[][]{\includegraphics[width=\linewidth]{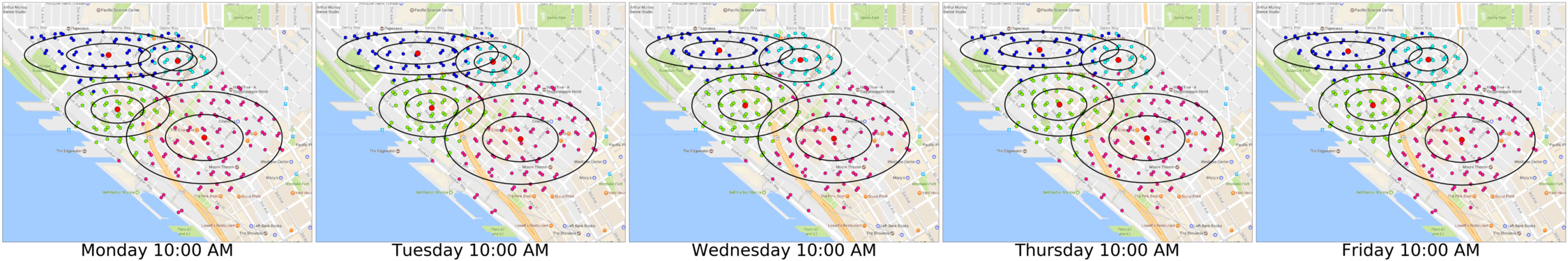}
    \label{fig:mixture_daily}}
    \newline
    \subfloat[][]{\includegraphics[width=\linewidth]{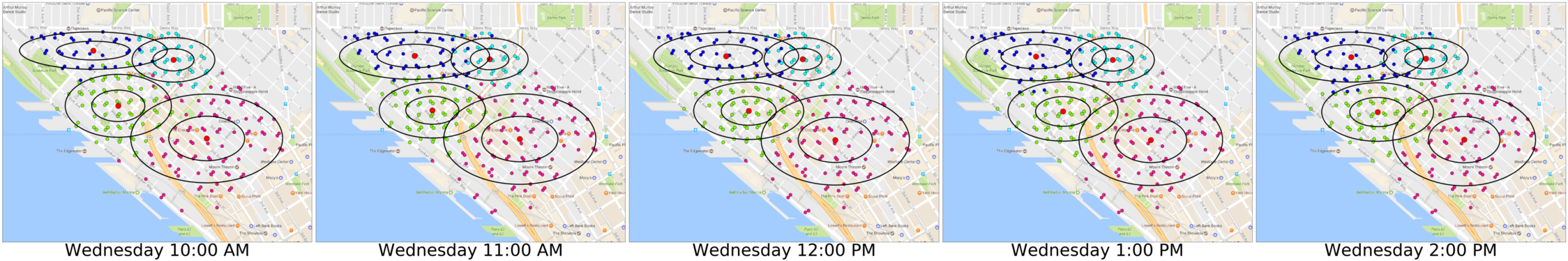}\label{fig:mixture_hourly}}
    \label{fig:mixture_many}
    \caption{The GMM learned in Belltown Monday--Friday from $8$AM--$4$PM are very similar. This figure shows a representative sample of this with Monday--Friday at $10$AM in Fig. \ref{fig:mixture_daily} and Wednesday $10$AM--$3$PM in Fig. \ref{fig:mixture_hourly} using the mean occupancies within Summer 2017.}
\end{figure*}


\subsection{Consistency of Parking Demand}
The results in Table \ref{table:results_summer} for Summer 2017 establish that spatial parking demand is consistent through time. With the exception of the first hour of paid parking in a day when occupancy is very low as drivers arrive, the mean consistency value at a hour of a day is very high ranging from $83.2\%$--$90.2\%$. Furthermore, on a given day of the week, even when including the less consistent first hour of the day, the mean consistency value is still very high ranging from $82.5\%$--$87.3\%$. In Table~\ref{table:results_mean} we present the mean daily consistency values for each season we analyze and observe the consistency values are nearly the same within each season. These results demonstrate that in the parking domain, making decisions based off historical trends is a reasonable thing to do, as behavior does not change considerably over time without changes to the system.

In addition to being able to quantify the consistency of demand in terms of block-faces belonging to distinct groups, we also investigate how the spatial centers of the mixture components change from week to week. We do this for a day of the week and hour of the day by using the $k$-means clustering
algorithm~\cite{Murphy:2012:MLP:2380985} on the centers of
components that were found at each date with the same corresponding day of the
week and hour of the day. Fig.~\ref{fig:centroids} shows an example of
clustering the centers from GMM learned on each Wednesday at $10$AM in
Summer 2017. The centers are tightly clustered with little change in location from week to week. 

By finding the centroids of each of the $k$-means clusters, and calculating the
mean distance from each centroid to the points in that respective cluster, we
can describe this change in terms of distance. In Fig.~\ref{fig:centroids}, we
find the average distance of the points to their respective centroids to be just
$36$ meters (m). The corresponding value at all other days of week and hours of the day
has mean of $71$m. In Summer 2016, Fall 2016, Winter 2017, Spring 2017 the results are comparable with values of $73$m, $73$m, $69$m, and $62$m respectively. This analysis has the advantage of being very interpretable in that we see the spatial centers are reliable in our model across time to within just a few street blocks.

\begin{table*}[t]
  \centering
  \caption{Spatial autocorrelation results for Belltown in each season for each
  method of creating the spatial weight matrix. The values indicate the
  percentage of instances in our data set---each instance given by the occupancy
  at a date and time---in which the $p$-value of Moran's $I$ was significant.
  The percentages in the first four rows use the methods of
  Section~\ref{local}, the fifth and sixth rows use the methods of
  Section~\ref{area}, the seventh and eighth rows use the methods of
  Section~\ref{mixture}, and the last row is the mean across the different
  season/year pairs we examined.}
  \label{table:results_auto}
  {\small      \begin{tabular}{|c||c|c|c|c|c|c|}
    \hline
 & \textbf{Summer/2016} & \textbf{Fall/2016} &
    \textbf{Winter/2017} & \textbf{Spring/2017} &
    \textbf{Summer/2017} & \textbf{Mean} \\
    \hline\hline
    $k=3$ &  57.4 &84.9 &85.0  &89.4 &  95.5 &    \textbf{82.4}    \\   \hline
    $k=5$ & 73.9 &93.3  &91.4 &95.3 &99.5 &\textbf{90.7}\\\hline
    $k=10$& 85.5 & 96.0 &96.7 &98.5 &99.4 &\textbf{95.2}\\\hline
    Distance & 31.3 &55.1 & 56.4 &44.7 &22.2 &\textbf{41.9}\\\hline\hline
    Area Connections & 53.1 & 68.1 &65.0&65.6 &59.8 & \textbf{62.3}\\\hline
    Area Distance & 66.3 &85.3 &84.0&85.2 &80.3 &\textbf{80.2}\\\hline\hline
    GMM Connections & 99.4 &  99.9 & 99.9&99.8  &100.0 &\textbf{99.8}\\\hline
    GMM Distance &  99.1 & 99.7 &99.9 &99.8 &100.0 &\textbf{99.7}\\\hline
  \end{tabular}
  }
\end{table*}

\subsection{Spatio-Temporal Insights}\label{insights}
One of the major insights we gain from the GMM approach is learning more about the time periods in which parking behavior is spatially similar. Notably, we find that Monday--Friday from 8AM--4PM nearly identical models are learned. Likewise, we find that Monday--Friday from 4PM--8PM very similar models are learned, which are different from those learned Monday--Friday from 8AM--4PM. We observe that models we learn for Saturday are quite unique and need to be considered on their own. Fig.~\ref{fig:mixture_daily} serves to show what we observe by considering a specific hour in the 8AM--4PM interval at each day of the week. Moreover, Fig.~\ref{fig:mixture_hourly} depicts using Wednesday that we see negligible change in the model from 8AM--4PM\footnote{We make animations available showing the model learned at each day of the week and hour of the day for each season we analyze at \href{https://github.com/fiezt/spatial-data-analysis/blob/master/animation/}{github.com/fiezt/spatial-data-analysis/blob/master/animation/}.}. To give an idea of what the model generally is like Monday--Friday from 4PM--8PM we refer the reader to Fig.~\ref{fig:mixture1}.

The preceding observations indicate that, based off of our model, it would make the most sense to have two weekday pricing periods---i.e.~8AM--4PM and 4PM--8PM---for the zones we commonly find at these respective time periods, and a unique Saturday pricing scheme. These results are compelling because they are quite different than the policies in place now, while still being surprisingly simple. Currently, the pricing periods in Belltown are from 8AM--11AM and 11AM--8PM with no individual consideration given to Saturdays. 

In comparing our model of spatial demand to the zones in place at present, we
identify some key similarities and differences. At nearly all paid parking times
during weekdays we learn a mixture component that covers a zone similar to that
of the South zone in Belltown. Yet in the North Zone of Belltown, our model
typically learns to divide up what is now the North Zone into three distinct
zones. This implies that the South Zone may in fact be simple enough to consider as is, while improvements can be made to policies in the North Zone.

\subsection{Spatial Autocorrelation Results}\label{auto_results}
Table~\ref{table:results_auto} gives the results of our spatial autocorrelation
analysis in Belltown for each of the seasons and methods of creating the spatial
weight matrix to evaluate particular questions of interest. The first four rows
give the results assessing the local and global spatial autocorrelation as
described in Section~\ref{local}, the fifth and sixth rows give the results
assessing the spatial autocorrelation in the current paid parking zones as
described in Section~\ref{area}, and the seventh and eighth rows give the results assessing the spatial autocorrelation in the GMM components as described in Section~\ref{mixture}.

The principal conclusions we draw from these results are as follows:
\begin{enumerate}
\item  The $k$-nearest neighbor method from Section~\ref{local} reveals that within Belltown there is almost always significant local spatial autocorrelation, indicating block-faces adjacent to each other see similar demand characteristics. The distance based method from Section~\ref{local}  confirms our previous observations that there is spatial heterogeneity in the demand globally within Belltown as the frequency of significant spatial autocorrelation is much lower.
\item The methods assessing the current paid parking zones from Section~\ref{area} demonstrate that while at times there is significant spatial autocorrelation within them, often this is not the case, attesting to the case that static pricing policies can be improved by considering location. 
\item The methods assessing the zones learned using the GMM from Section~\ref{mixture} demonstrate that our approach provides meaningful improvement. The results show that at nearly all times the spatial autocorrelation within the mixture components is significant. 
\end{enumerate}

The spatial autocorrelation results give a way to quantify the improvement of
our GMM approach over the current paid parking zones. The importance of spatial
autocorrelation in this context is that in the design of static policies, a
policy that is applied uniformly over a zone will be more effective if the
demand characteristics are similar within the zone. In addition, if attempting to influence behavior through control methods, specifically targeted information and incentive schemes, it allows the designer to consider zones as groups of users with similar preferences.

One intrinsic property of our model that may guide the improvement is the difference in the number of zones we consider using our model selection criteria. This is a simple way to avoid heuristics while maintaining desired simplicity since the increase in zones is modest.

The zones we find also have decidedly lower variance in the demand than the current paid parking zones in Belltown. In the seasons we analyze the variance in the occupancy in the GMM zones averaged over all paid parking times has a minimum of $5.8\%$, mean of $6.0\%$, and a maximum of $6.2\%$. The corresponding variance in the occupancy in the paid parking zones has a minimum of $9.0\%$, mean of $9.5\%$, and a maximum of $10.1\%$.

\begin{figure}[t!]
    \centering
    \subfloat[][Friday 7PM in June, 2016]{\includegraphics[width=0.8\columnwidth]{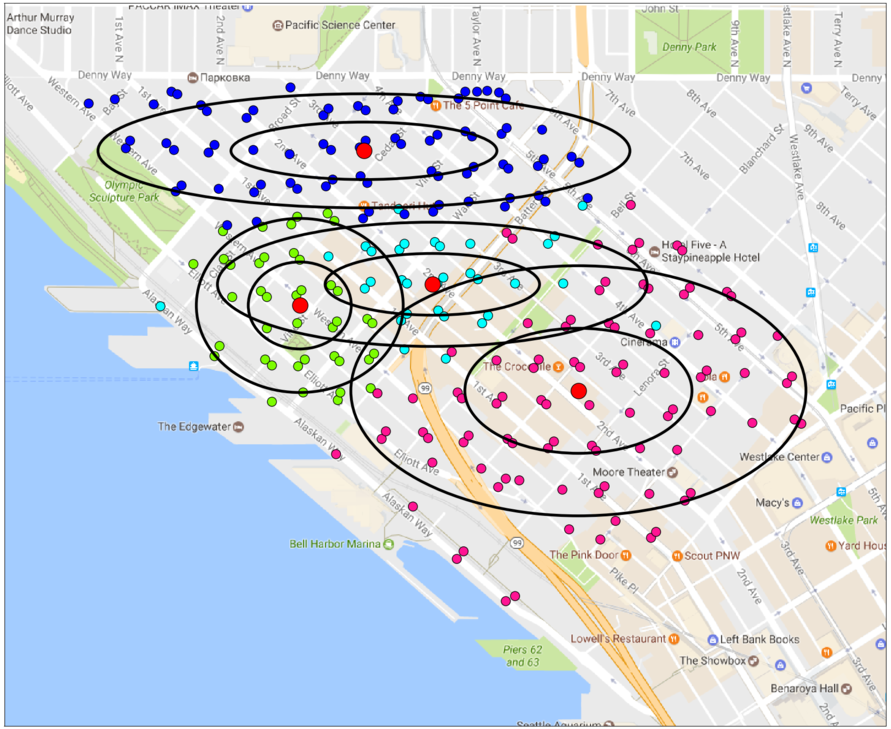}\label{fig:mixture_pre}}
   
    \subfloat[][Friday 7PM in June, 2017]{\includegraphics[width=0.8\columnwidth]{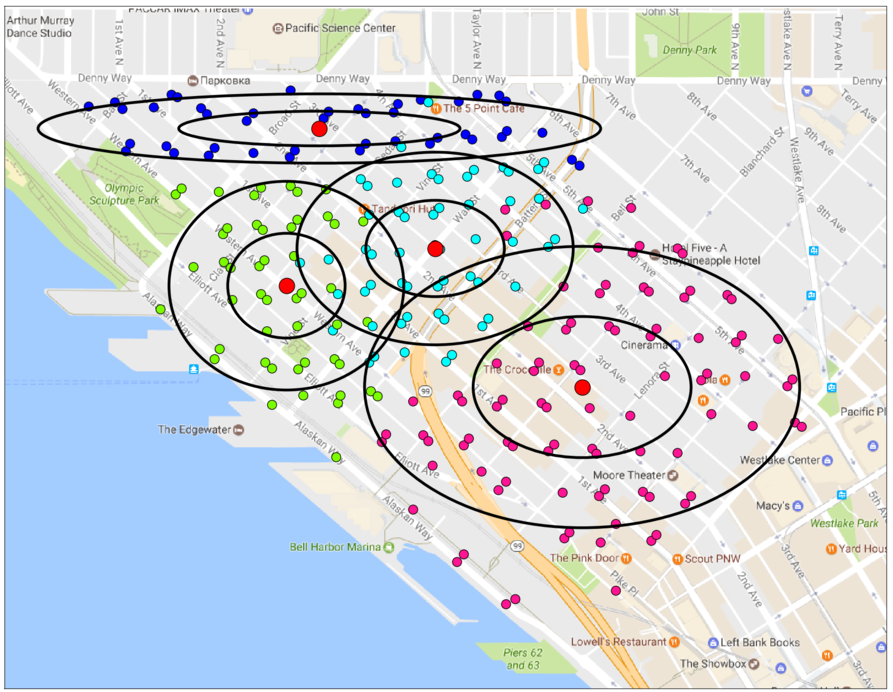}\label{fig:mixture_post}}
    \caption{GMM analysis before (Fig.~\ref{fig:mixture_pre}) and after (Fig.~\ref{fig:mixture_post}) the price change in Belltown using the mean occupancies in the respective months.}
    \label{fig:mixture_price}
\end{figure}
 
\begin{figure*}[t!]
    \centering
    \subfloat[][Belltown and Commercial Core]{\includegraphics[width=4.4cm, height=3.7cm]{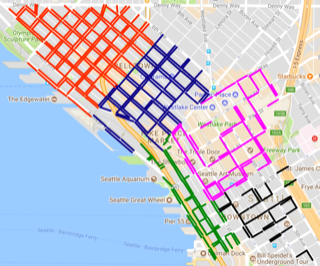}\label{fig:belltown_commcore}}\hfill
    \subfloat[][Tuesday $11$AM]{\includegraphics[width=4.4cm, height=3.7cm]{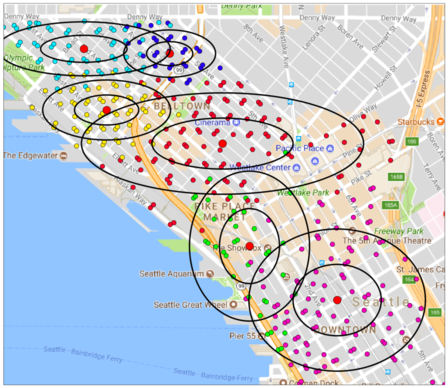}\label{fig:mixture_belltown_commcore}}\hfill
    \subfloat[][Belltown and Denny]{\includegraphics[width=4.4cm, height=3.7cm]{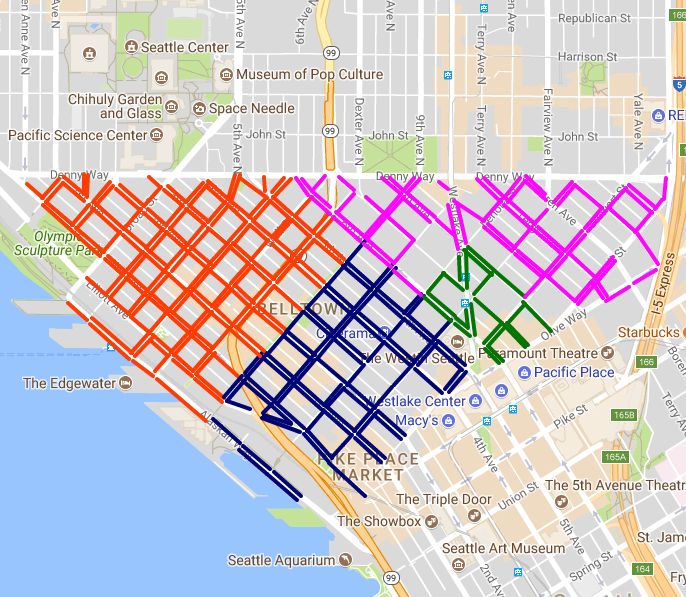}\label{fig:belltown_denny}}\hfill
    \subfloat[][Tuesday $1$PM]{\includegraphics[width=4.4cm, height=3.7cm]{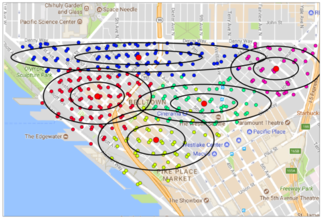}\label{fig:mixture_belltown_denny}}
    \caption{GMM analysis combining Belltown with Commercial Core and Belltown
    with Denny. (a) Map of the paid parking zones in Belltown (red, blue) and
Commerical Core (pink, green, black). (b) GMM with six components learned using
the mean occupancies in Belltown and Commerical Core within Summer 2017 at Tuesday $11$AM. (c) Map of the paid parking zones in Belltown (red, blue) and Denny (pink, green). (d) GMM with five components learned using the mean
occupancies in Belltown and Denny within Summer 2017 at Tuesday $1$PM.}
    \label{fig:multiple_neighborhoods}
\end{figure*}

\begin{table*}[t]
  \centering
  \caption{Mean daily consistency metric results for each season we analyze considering Belltown combined with Denny and Belltown combined with Commercial Core. Units are in percentage.}
  \label{table:results_multiple_consist}
  {\small      \begin{tabular}{|c||c|c|c|c|c|}
    \hline
    & \textbf{Summer/2016 }& \textbf{Fall/2016} & \textbf{Winter/2017} &
    \textbf{Spring/2017} & \textbf{Summer/2017}  \\
    \hline\hline
    Belltown \& Denny & 80.9 & 81.9 & 83.0 & 83.0 & 82.8 \\ 
    \hline
    Belltown \& Commercial Core    & 75.7 & 78.7 & 78.5 & 79.8 & 76.5 \\ 
    \hline
\end{tabular}}
\end{table*}

\begin{table}[t]
  \centering
  \caption{Spatial autocorrelation results averaged over each season/year we
  analyze for each method of creating the spatial weight matrix described in
  Section~\ref{spatial_autocorrelation} considering Belltown combined with Denny and Belltown combined with Commercial Core. The first four rows use the methods of Section~\ref{local}, the fifth and sixth rows use the methods of Section~\ref{area}, the seventh and eighth rows use the methods of
  Section~\ref{mixture}. The values are the percentage of instances that the $p$-value of Moran's $I$ was significant.}
  \label{table:results_multiple_auto}
  {\small     \begin{tabular}{|l||c|c|}
      \hline \textbf{Belltown w/}& \textbf{Denny}  & \textbf{Commercial
          Core} \\\hline\hline
      $k$ = 3 & 1.7 &      14.0  \\\hline  
      $k$ = 5 &1.0 & 10.9\\\hline
      $k$ = 10 & 1.7 &9.0\\\hline
      Distance &  1.4 &9.0\\\hline\hline
      Area Connections &  89.1 &88.0\\\hline
      Area Distance & 88.8 &87.6\\\hline\hline
      GMM Connections & 98.4 & 99.9\\\hline
      GMM Distance &  98.3 &  99.8 \\
    \hline
\end{tabular}}
\end{table}

\subsection{Seasonal and Price Changes}
In previous sections, we have seen that the consistency, spatial
autocorrelation, and variance results in Belltown are similar between seasons.
Likewise we find that the GMM we learn within the different seasons coincide.
These results indicate that in terms of the spatial demand, the variation
between seasons is insignificant, despite that in Section~\ref{season} we showed the variation in occupancy is non-negligible. Thus, we conclude that demand does indeed fluctuate over time, but the fluctuation in the relative levels of demand between block-faces is inconsequential, implying static policies can be robust to seasonal variation effects.

Analyzing the price change in 2016 in Belltown we draw similar conclusions.
While occupancy decreased from the month before the price change to the month
one year following the price change, the models we learn using data from the two
time periods hardly change. Fig.~\ref{fig:mixture_price} shows an example of
this. These results corroborate work investigating the SFPark study  suggesting price control methods may not produce the desired changes in behavior~\cite{pierce2013getting}.

\subsection{Going Beyond Belltown}
Up until this point, we have focused on the Belltown neighborhood alone. By examining neighborhoods together, we assess the fact that current paid parking zones are set based on existing neighborhood boundaries. This is an important matter because there is no particular reason that pricing zones should be set within existing neighborhood boundaries. In bypassing this requirement we can determine whether neighborhood boundaries should carry any weight.

We consider the Belltown neighborhood together with the Commercial Core
neighborhood and with the Denny neighborhood, both of which are adjacent to
Belltown. In Fig.~\ref{fig:multiple_neighborhoods}, we present examples using
data from Summer 2017 of the GMM approach as well as the current paid
parking zones. Like in the case of considering Belltown alone, we find a
different number of zones using our model selection criteria than are in place
now. In the case of Belltown and Commercial Core, we find six mixture components
with five zones designated, and in the case of Belltown and Denny, we find five
mixture components with four zones designated. In the examples in
Fig.~\ref{fig:mixture_belltown_commcore} and Fig.~\ref{fig:belltown_commcore}
the zones from the learned GMM cross many of the existing paid area boundaries
as well as the neighborhood boundaries, confirming that it is worth looking
beyond arbitrary existing boundaries. In these cases however, the zones we learn are closer to the designated zones by the city of Seattle than when we
considered Belltown on its own.

The spatial autocorrelation results given in Table~\ref{table:results_multiple_auto} reflect this fact, as the frequency of
significant spatial autocorrelation in the designated paid parking zones is
higher than was the case when only looking in Belltown. This says that what is
in place in Commercial Core and Denny at present, may be reasonable since we
learn something similar. The intriguing characteristic of these examples is that when adding the new neighborhoods the local autocorrelation is generally not significant. This can be attributed to these particular neighborhoods, Commercial Core and Denny, since within Belltown it was the case. Despite this matter, the results within the GMM zones remain satisfactory, suggesting that our model can discover acceptable groups of block-faces in the more challenging case where blocks within local neighborhoods may not see as similar of demand characteristics.

The consistency of spatial demand results in Table~\ref{table:results_multiple_consist} are close to those as when we only considered Belltown, albeit slightly lower, possibly owing to there being more flexibility in modeling with more components. There is also comparable temporal characteristics, in that the spatial autocorrelation, consistencies, and GMM we learn were nearly the same between seasons.

%% file: sections/discussion.tex
We provide an in depth analysis of the spatio-temporal characteristics of
parking demand using real data, as well as an interpretable way to find zones
where there is spatial homogeneity using a GMM. The work has the potential to
allow for more informed decision-making in both policy decisions as well as in designing targeted information and incentive schemes. Furthermore, we establish that spatial parking demand is consistent, which is to say that without changes to management or infrastructure, learned models will hold up over time. While we focus on Seattle, our methods leverage a now common data source of paid parking transactions from cities, making the models and analysis we use flexible enough to be applied in many other communities.

Parking management and policy is a difficult problem because of the many competing needs that need to be balanced. A potentially viable and under-explored approach to the problem is coupling new pricing schemes with targeted ads and incentives. Through our collaboration with SDOT, we seek to use this work to identify zones of similar demand in support of designing such schemes. 

Towards this end, we are investigating a multi-arm bandit framework to learning
user response while matching information and incentives in GMM identified zones. In this problem we consider each zone as a group of users with similar preferences which are evolving over time, dependent on how information or incentives are matched. We aim to solve the combinatorial bandit problem of learning how to optimally match sets of information and incentives to sets of location zones in a computationally efficient manner by taking advantage of correlations between dependent arms. In addition to theoretical analysis of the resulting bandit algorithms, we plan to implement developed strategies in a living lab setting in Seattle neighborhoods with the aid of SDOT.

%% file: 2017IEEEITS.bbl
\begin{thebibliography}{10}
\providecommand{\url}[1]{#1}
\csname url@samestyle\endcsname
\providecommand{\newblock}{\relax}
\providecommand{\bibinfo}[2]{#2}
\providecommand{\BIBentrySTDinterwordspacing}{\spaceskip=0pt\relax}
\providecommand{\BIBentryALTinterwordstretchfactor}{4}
\providecommand{\BIBentryALTinterwordspacing}{\spaceskip=\fontdimen2\font plus
\BIBentryALTinterwordstretchfactor\fontdimen3\font minus
  \fontdimen4\font\relax}
\providecommand{\BIBforeignlanguage}[2]{{%
\expandafter\ifx\csname l@#1\endcsname\relax
\typeout{** WARNING: IEEEtran.bst: No hyphenation pattern has been}%
\typeout{** loaded for the language `#1'. Using the pattern for}%
\typeout{** the default language instead.}%
\else
\language=\csname l@#1\endcsname
\fi
#2}}
\providecommand{\BIBdecl}{\relax}
\BIBdecl

\bibitem{shoup:2005aa}
D.~Shoup, \emph{The High Cost of Free Parking}.\hskip 1em plus 0.5em minus
  0.4em\relax American Planning Association, 2005.

\bibitem{shoup2007gone}
D.~Shoup and H.~Campbell, ``Gone parkin','' \emph{The New York Times}, vol.~29,
  2007.

\bibitem{levy2010evaluation}
J.~I. Levy, J.~J. Buonocore, and K.~Von~Stackelberg, ``Evaluation of the public
  health impacts of traffic congestion: a health risk assessment,''
  \emph{Environmental health}, vol.~9, no.~1, p.~65, 2010.

\bibitem{zhang2013air}
K.~Zhang and S.~Batterman, ``Air pollution and health risks due to vehicle
  traffic,'' \emph{Science of the total Environment}, vol. 450, pp. 307--316,
  2013.

\bibitem{dowling2017much}
C.~Dowling, T.~Fiez, L.~Ratliff, and B.~Zhang, ``How much urban traffic is
  searching for parking?'' \emph{arXiv preprint arXiv:1702.06156}, 2017.

\bibitem{geng2013new}
Y.~Geng and C.~G. Cassandras, ``New “smart parking” system based on
  resource allocation and reservations,'' \emph{IEEE Transactions on
  Intelligent Transportation Systems}, vol.~14, no.~3, pp. 1129--1139, 2013.

\bibitem{griggs2016design}
W.~Griggs, J.~Y. Yu, F.~Wirth, F.~H{\"a}usler, and R.~Shorten, ``On the design
  of campus parking systems with qos guarantees,'' \emph{IEEE Transactions on
  Intelligent Transportation Systems}, vol.~17, no.~5, pp. 1428--1437, 2016.

\bibitem{kotb2016iparker}
A.~O. Kotb, Y.-C. Shen, X.~Zhu, and Y.~Huang, ``{iParker-—A New Smart
  Car-Parking System Based on Dynamic Resource Allocation and Pricing},''
  \emph{IEEE Transactions on Intelligent Transportation Systems}, vol.~17,
  no.~9, pp. 2637--2647, 2016.

\bibitem{sun2016discriminated}
D.~J. Sun, X.-Y. Ni, and L.-H. Zhang, ``A discriminated release strategy for
  parking variable message sign display problem using agent-based simulation,''
  \emph{IEEE Transactions on Intelligent Transportation Systems}, vol.~17,
  no.~1, pp. 38--47, 2016.

\bibitem{bagula2015design}
A.~Bagula, L.~Castelli, and M.~Zennaro, ``On the design of smart parking
  networks in the smart cities: An optimal sensor placement model,''
  \emph{Sensors}, vol.~15, no.~7, pp. 15\,443--15\,467, 2015.

\bibitem{qian2012optimal}
Z.~Qian and R.~Rajagopal, ``Optimal dynamic pricing for morning commute parking
  with occupancy information,'' \emph{Transportation Research Part B}, 2012.

\bibitem{qian:2012aa}
Z.~S. Qian, F.~E. Xiao, and H.~Zhang, ``Managing morning commute traffic with
  parking,'' \emph{Transportation Research Part B: Methodological}, vol.~46,
  no.~7, pp. 894--916, 2012.

\bibitem{zoeter2014new}
O.~Zoeter, C.~Dance, S.~Clinchant, and J.-M. Andreoli, ``New algorithms for
  parking demand management and a city-scale deployment,'' in \emph{{Proc. 20th
  ACM SIGKDD Inter. Conf. Knowledge Discovery and Data Mining}}.\hskip 1em plus
  0.5em minus 0.4em\relax ACM, 2014, pp. 1819--1828.

\bibitem{dowling2017optimizing}
C.~Dowling, T.~Fiez, L.~Ratliff, and B.~Zhang, ``Optimizing curbside parking
  resources subject to congestion constraints,'' in \emph{Proc. 56th IEEE Conf.
  Decision and Control (arXiv:1703.07802)}, 2017.

\bibitem{pierce2013getting}
G.~Pierce and D.~Shoup, ``Getting the prices right: an evaluation of pricing
  parking by demand in san francisco,'' \emph{J. American Planning
  Association}, vol.~79, no.~1, pp. 67--81, 2013.

\bibitem{ratliff:2016aa}
L.~J. Ratliff, C.~Dowling, E.~Mazumdar, and B.~Zhang, ``To observe or not to
  observe: Queuing game framework for urban parking,'' in \emph{Proc.~55th
  Conf.~Decision and Control}, 2016, pp. 5286--5291.

\bibitem{yang2017turning}
S.~Yang and Z.~S. Qian, ``Turning meter transactions data into occupancy and
  payment behavioral information for on-street parking,'' \emph{Transportation
  Research Part C: Emerging Technologies}, vol.~78, pp. 165--182, 2017.

\bibitem{lin2017survey}
T.~Lin, H.~Rivano, and F.~Le~Mou{\"e}l, ``A survey of smart parking
  solutions,'' \emph{IEEE Transactions on Intelligent Transportation Systems},
  2017.

\bibitem{glasnapp2014understanding}
J.~Glasnapp, H.~Du, C.~Dance, S.~Clinchant, A.~Pudlin, D.~Mitchell, and
  O.~Zoeter, ``Understanding dynamic pricing for parking in los angeles: Survey
  and ethnographic results,'' in \emph{Inter. Conf. HCI in Business}.\hskip 1em
  plus 0.5em minus 0.4em\relax Springer, 2014, pp. 316--327.

\bibitem{ma2013parking}
X.~Ma, X.~Sun, Y.~He, and Y.~Chen, ``Parking choice behavior investigation: A
  case study at beijing lama temple,'' \emph{Procedia-Social and Behavioral
  Sciences}, vol.~96, pp. 2635--2642, 2013.

\bibitem{carney2013bringing}
\BIBentryALTinterwordspacing
N.~Carney. (2013) Bringing markets to meters. [Online]. Available:
  \url{http://datasmart.ash.harvard.edu/news/article/bringing-markets-to-meters-312}
\BIBentrySTDinterwordspacing

\bibitem{fiez:2017aa}
T.~Fiez, L.~J. Ratliff, C.~Dowling, and B.~Zhang, ``Data-driven spatio-temporal
  modeling of parking demand,'' in \emph{submitted to ACC}, 2017.

\bibitem{statatlas}
\BIBentryALTinterwordspacing
StatisticalAtlas. (2017) Population by neighborhood in seattle. [Online].
  Available:
  \url{https://statisticalatlas.com/neighborhood/Washington/Seattle/Belltown/Population}
\BIBentrySTDinterwordspacing

\bibitem{dempster1977maximum}
A.~P. Dempster, N.~M. Laird, and D.~B. Rubin, ``Maximum likelihood from
  incomplete data via the em algorithm,'' \emph{J. Royal Statistical Society.
  Series B (Methodological)}, pp. 1--38, 1977.

\bibitem{wu1983convergence}
C.~J. Wu, ``On the convergence properties of the em algorithm,'' \emph{The
  Annals of statistics}, pp. 95--103, 1983.

\bibitem{boyles1983convergence}
R.~A. Boyles, ``On the convergence of the em algorithm,'' \emph{J. Royal
  Statistical Society. Series B (Methodological)}, pp. 47--50, 1983.

\bibitem{Murphy:2012:MLP:2380985}
K.~P. Murphy, \emph{Machine Learning: A Probabilistic Perspective}.\hskip 1em
  plus 0.5em minus 0.4em\relax The MIT Press, 2012.

\bibitem{schwarz1978estimating}
G.~Schwarz \emph{et~al.}, ``Estimating the dimension of a model,'' \emph{The
  annals of statistics}, vol.~6, no.~2, pp. 461--464, 1978.

\bibitem{moran1950notes}
P.~A. Moran, ``Notes on continuous stochastic phenomena,'' \emph{Biometrika},
  vol.~37, no. 1/2, pp. 17--23, 1950.

\end{thebibliography}
